\documentclass[twocolumn,amsmath,showkeys,showpacs,prb]{revtex4}
\pdfoutput=1
\usepackage{dcolumn}
\usepackage{bm}
\usepackage[T1]{fontenc}
\usepackage{amsmath}
\usepackage{graphicx}

\usepackage{hyperref}
\hypersetup{colorlinks=true, linkcolor=blue, citecolor=red, urlcolor=magenta, pdftitle={Ferroelectric and antiferroelectric instabilities in BaTiO3/BaO superlattices from first-principles}, pdfauthor={Eric Bousquet, Javier Junquera and Philippe Ghosez}}

\begin{document}
\title{First-principles study of competing ferroelectric and antiferroelectric instabilities in 
       BaTiO${_3}$/BaO superlattices}

\author{Eric Bousquet$^{1,2}$}
\author{Javier Junquera$^3$} 
\author{Philippe Ghosez$^1$}
\affiliation { $^1$Institut de physique (B5), Universit\'e de Li\`ege,  
              B-4000 Sart Tilman, Belgium}
\affiliation{ $^2$Materials Department, University of California, 
             Santa Barbara, CA 93106, USA}
\affiliation{ $^3$Departamento de Ciencias de la Tierra y F\'{\i}sica
             de la Materia Condensada, Universidad de Cantabria,
             E-39005 Santander, Spain}

\begin{abstract}
We report a first-principles study of (BaTiO${_3}$)$_m$/(BaO)$_n$ superlattices 
for a wide range of periodicities $m/n$.  We show that such a system develops a 
polar zone-center instability for sufficiently large \textit{m}/\textit{n} ratio, which can be
understood, at least qualitatively, from a simple electrostatic model and should 
lead to a ferroelectric ground-state.  However, the analysis of the phonon dispersion 
curves also points out the appearance of stronger antiferroelectric instabilities 
at the zone boundaries around $m=4$, before the critical ratio for ferroelectricity is 
reached and which still dominate beyond it. The dominant character of the anti-ferroelectric
instability is explained from the depolarizing field which hardens the ferroelectric
mode. This analysis allows us to predict that, (BaTiO${_3}$)$_m$/(BaO)$_n$ superlattices 
should present an antiferroelectric ground state for $m$ larger than 4, which should 
smoothly evolve to a multidomain structure for increasing $m$ values and only become 
ferroelectric for large $m$.
\end{abstract}


\maketitle

\section{Introduction}

During the recent years, numerous works have been devoted to the study of 
interfacial effects in ferroelectric 
nanostructures.\cite{ghosez2006,dawber2005b} 
Both theoretical and experimental works have focused on different types of 
ferroelectric multilayers and superlattices including 
(i) ferroelectric ultrathin films between metalic electrodes 
(FE/Me interface),\cite{junquera2003,stengel2006,aguado2008,stengel2009,Stengel-09.c} 
(ii) superlattices combining two ferroelectric materials 
(FE/FE interface),\cite{qu1997,george2001,huang2003} and 
(iii) superlattices in which a ferroelectric alternates with an incipient 
ferroelectric (FE/iFE interface).\cite{neaton2003,johnston2005,dawber2005,bousquet2008,sepliarsky2001} 
These studies highlighted the fact that three main factors govern the physics 
of multilayers: epitaxial strain, electrical boundary conditions and 
interfacial effects.

The case of ferroelectric/insulator superlattices (FE/I interface) 
in which a ferroelectric alternates with a regular insulator was only 
marginally addressed in the literature. This might be due to the fact that 
no enhancement of the ferroelectric properties is expected in such systems 
due to the detrimental effect  of the regular insulator on the ferroelectric 
properties. For some time, only the interface between AO and ABO$_3$ oxides,
where A is an alkaline-earth atom such as Sr or Ba, attracted some 
interest since AO oxides are playing a key role as buffer layer in the 
epitaxial growth of perovskites directly on a 
silicon substrate.\cite{mckee1998}
A first-principles study was previously reported by some of us for 
(BaTiO$_3$)$_m$/(BaO)$_n$ and (SrTiO$_3$)$_m$/(SrO)$_n$ superlattices
grown on Si,\cite{junquera2003b} but only the paraelectric periodicity 
$n$=6 and $m$=5 was considered and the ferroelectric properties were not 
explicitly discussed.

Nevertheless the focus on these particular FE/I interfaces has lately
increased. The Ruddlesden-Poper series can be considered as an example 
of this type.\cite{Ruddlesden,Nakhmanson2008,nakhmanson2010} This renewed interest is fueled
by the possibility of modifying in a controlled way the delicate
balance between all the instabilities present at the bulk ABO$_{3}$
perovskites, that might be ferroelectric (FE), antiferroelectric (AFE),
and/or antiferrodistortive (AFD) phase transitions by changing
the composition and/or periodicity of the series.
The perspectives of stabilizing new exotic phases not present at the bulk 
parent compounds is a strong driving force for
these studies.\cite{bousquet2008}

In this paper, we study the case of (BaTiO$_3$)$_m$/(BaO)$_n$ 
superlattices epitaxially grown on a SrTiO$_3$ substrate as a prototypical 
example of FE/I superlattices. 
We first consider the predictions of a simple electrostatic model as 
reported in Ref.~\onlinecite{dawber2005}. Then, we present 
first-principles calculations of the phonon band structure of the 
paraelectric (BaTiO$_3$)$_m$/(BaO)$_n$ superlattices for various 
layer thicknesses $m$ and $n$, and discuss the effects 
of the periodicity on the vibrational properties. 
We show that the predicted ground state 
differs from that of the simple electrostatic model,
and explain the reason for the discrepancy.

\section{Technical details}

First-principles simulations were performed in the framework of the 
Density Functional Theory (DFT) as implemented in the 
{\sc Abinit} package.\cite{abinit} All the results were calculated 
through the local density approximation (LDA) for the 
exchange-correlation energy. 
An energy cut-off of 45 Ha was used for the planewave expansion. 
The Teter parametrization~\cite{teter1993} for pseudopotentials was employed
where the following orbitals were treated as valence states: 
5$s$, 5$p$ and 6$s$ for the Ba atom, 3$s$, 3$p$, 3$d$ and 4$s$ for the Ti atom
and 2$s$ and 2$p$ for the O atom. From the smallest to the biggest 
superlattices, tetragonal Monkhorst-Pack meshes from respectively 
$6 \times 6 \times 3$ to $6 \times 6 \times 1$ were considered for the 
Brillouin zone sampling in order to produce accurate results.
The vibrational properties and Born effective charges were calculated 
using the Density functional perturbation theory (DFPT).\cite{gonze1997}

BaTiO$_3$ can be epitaxially grown on BaO.\cite{mckee1991} 
The epitaxy is such that 
ABO$_3$(001)$\parallel$AO(001) and ABO$_3<$110$>\parallel$AO\-$<$100$>$, 
which means that the BaTiO$_3$ unit cell is rotated $45^\circ$  around 
the (001) BaO direction. 

The superlattices were simulated by means
of a supercell approximation with periodic boundary
conditions, so short-circuit electrostatic
boundary conditions across the whole supercell are naturally impossed.
To define the cell that is periodically repeated in space,
a generic formula (BaO-TiO$_2$)$_m$/(BaO)$_n$ was used and
labeled by $m/n$, 
where $n$ is the number of BaO oxide atomic layer and $m$ the number 
of BaTiO$_3$ formula units.\cite{junquera2003b, mckee2001}

The epitaxial strain was treated by fixing the cubic in-plane lattice constant
of the superlattices. Only the case of an epitaxial strain corresponding to an
hypothetical SrTiO$_3$ substrate was considered (theoretical LDA relaxed cubic
cell parameter of SrTiO$_3$: $a$ = 3.84 \AA). For each thickness, 
we performed structural optimization of the atomic positions and 
the out-of-plane cell parameter of the superlattice in its highly 
symmetric phase corresponding
to the tetragonal space group \textit{P4/mmm} (No 123).

To establish the notation, we will call the plane parallel to the interface 
the ($x,y$) plane, whereas the perpendicular direction will be referred to 
as the $z$ axis.

\section{Electrostatic model}

As explained in Ref.~\onlinecite{ghosez2006,dawber2005}, 
the energy of a ferroelectric superlattice between materials 1 and 2 
under short-circuit boundary conditions
can be estimated within a simple model 
which assumes homogeneous polarization in each layer and neglect 
interface corrections, as the sum of the internal energy for the 
two compounds forming the superlattice and an electrostatic correction,

\begin{equation}
   E (P_1,P_2) =  m U_1(P_1) + n U_2(P_2) + E_{elec}(m,n,P_1,P_2).
   \label{Etot1}
\end{equation}

\noindent In first approximation, the internal energies $U_1$ and $U_2$ 
can be determined at the bulk level in the absence of a macroscopic 
electric field, while the term $E_{elec}$ takes into 
account the electrostatic energy cost related to the polarization mismatch 
between the two types of layers.
From now on, we will assume that material 1 is BaTiO$_{3}$ (BTO)
and material 2 is BaO (BO).

\subsection{Bulk internal energies}

In their relaxed cubic structures, we obtained the lattice constants 
$a_{{\rm BO}}$=5.46 \AA\ and $a_{{\rm BTO}}$=3.94 \AA\ 
for BaO and BaTiO$_3$ respectively, 
which are in good agreement with previous LDA results.\cite{junquera2003b}
In the NaCl structure, a cell parameter of 5.46 \AA\ in the BaO bulk 
corresponds to a Ba-Ba distance of 3.86 \AA.
Therefore, when grown on a SrTiO$_3$ substrate, assuming a theoretical 
cubic in-plane lattice constant of 3.84 \AA, BaTiO$_3$ is under 
compressive strain and becomes tetragonal, as well as BaO 
(the epitaxial strain being applied on the Ba-Ba distance) 
in a smaller extent.\cite{strainbo}
After relaxation with this epitaxial constraint, the out-of-plane cell 
parameters of the paraelectric references are
$c^0_{{\rm BO}}$=5.48 \AA\ and $c^0_{{\rm BTO}}$=4.01 \AA\ 
for BaO and BaTiO$_3$ respectively (where the superscript ``0'' refers to 
the fact that it is the $c$-parameter at zero polarization). The associated 
Born effective charge and electronic
dielectric tensors are reported in Table~\ref{ZepsBulk}. 
These values are comparable to those in the cubic structure, with a
small anisotropy due to the tetragonal symmetry: most components of the 
Born effective charges and electronic dielectric tensors are smaller in
the $z$ than in the $x/y$ directions. 
The frequency of the lowest transverse optic (TO)
mode are also reported in Table~\ref{ZepsBulk} together with
their mode effective charges. In this constrained paraelectric configuration, 
BaO is stable but the mode polarized along the $z$ direction has a
slightly lower frequency than those along the $x/y$ 
directions.\cite{bousquet2010} BaTiO$_3$ shows a ferroelectric instability 
along the $z$ direction with an imaginary frequency of 180i cm$^{-1}$
and an associated mode effective charge of 8.47 e while in the $x/y$ 
directions, the frequency of the soft mode is shifted to
real frequency (195 cm$^{-1}$) and its mode effective charge 
is slightly reduced (8.00 e).

\begin{table}[htbp!]
   \begin{center}
      \begin{tabular}{l r@{.}l r@{.}l r@{.}l r@{.}l}
         \hline
         \hline
                                       & 
         \multicolumn{4}{c}{BaO}       & 
         \multicolumn{4}{c}{BaTiO$_3$} \\
                                       &
         \multicolumn{2}{c}{\ $x/y$}   &
         \multicolumn{2}{c}{$z$}       & 
         \multicolumn{2}{l}{\ $x/y$}   &
         \multicolumn{2}{c}{$z$}       \\
         \hline
         $a$                             &  
         3&84                            & 
         5&48                            &  
         3&84                            & 
         4&01                            \\
         $Z^{\ast}$(Ba)                  &  
         2&83                            & 
         2&78                            &  
         2&81                            & 
         2&77                            \\
         $Z^{\ast}$(O$_1$)               &
         -2&83                           &
         -2&78                           &
         -2&27                           &
         -5&29                           \\
         $Z^{\ast}$(Ti)                  &
         \multicolumn{2}{c}{-}           & 
         \multicolumn{2}{c}{-}           &  
         7&48                            & 
         6&80                            \\
         $Z^{\ast}$(O$_{2/3}$)           &
         \multicolumn{2}{c}{-}           & 
         \multicolumn{2}{c}{-}           &
         \multicolumn{2}{l}{-2.13/-5.90} &
         -2&14                           \\
         $\epsilon^{\infty}$             &
         4&39                            & 
         4&38                            &  
         6&65                            & 
         6&45                            \\
         $\omega_{TO}$                   & 
         \multicolumn{2}{c}{118}         &
         \multicolumn{2}{c}{74}          & 
         \multicolumn{2}{l}{\ 195}       &  
         \multicolumn{2}{c}{\ 180i}      \\
         \={Z}$^{\ast}$                  &  
         3&14                            & 
         3&08                            &  
         8&00                            & 
         8&47                            \\
         \hline
         \hline
      \end{tabular}
      \caption{ Cell parameters ($a$, in \AA), 
                diagonal components component of the Born effective charge tensors 
                (Z$^{\ast}$, in e), 
                diagonal components component of the optical dielectric constant tensors
                ($\epsilon^{\infty}$),
                frequency of the lowest TO modes ($\omega_{TO}$, in cm$^{-1}$) 
                and associated mode effective charges (\={Z}$^{\ast}$, in e) 
                for the bulk paraelectric
                BaO and BaTiO$_3$, under the constraint of an in-plane
                lattice constant imposed by an hypothetical SrTiO$_3$ 
                substrate (3.84 \AA).
                O$_{1}$ refers to the O in the BaO plane while 
                O$_{2/3}$ stands for the two equivalent oxygens in 
                the TiO$_{2}$ planes.The diagonal components of $Z^{\star}$ and 
                $\epsilon^{\infty}$ along $x$ and $y$ 
                are equal by symmetry.
               }
      \label{ZepsBulk}
   \end{center}
\end{table}

\begin{center}
   \begin{figure}[htbp!]
      \begin{center}
         \includegraphics[width=7.5cm, height=4.29cm, angle=0]{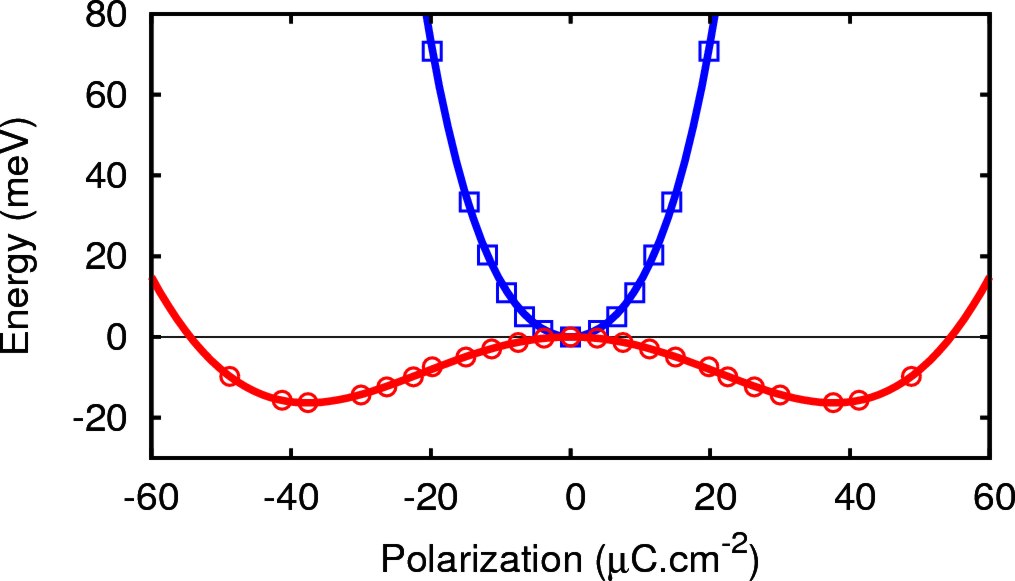}
      \end{center}
      \caption {(Color online) Energy as a function of 
                spontaneous polarization for epitaxial bulks.
                The blue curve with squares corresponds to the tetragonal bulk
                BaO and the red curve with circles is the double well of 
                tetragonal BaTiO$_3$ bulk.
                For each single point energy, the out-of-plane cell parameter 
                was relaxed.}
      \label{wells}
   \end{figure}
\end{center}

The compressive epitaxial strain imposed by the substrate favors a 
ferroelectric $c$-phase~\cite{dieguez2005} and, therefore,
BaTiO$_{3}$ effectively becomes an uniaxial ferroelectric.
In what follows, we shall assume that all the polarization
and electric fields are directed along $z$. 
A full relaxation of both the atomic coordinates 
and lattice vectors gives rise to a 
ferroelectric ground state with a spontaneous polarization of 
38 $\mu$C/cm$^{2}$ and a relaxed out-of-plane cell parameter of 4.07 \AA. 
The difference of energy between the ferroelectric ground state and the
paraelectric reference is $\Delta E$=16 meV which corresponds to the 
depth of the double well associated to the evolution of the potential
energy with the polarization (see red curve in Fig.~\ref{wells}). 
In principle, the determination of the shape of the double well 
potential energy, $U(P)$, would require calculation at
constrained $P$ as proposed in Ref.~\onlinecite{sai2002}.
Here we used a more approximate but still reasonable approach. 
First, we froze in different fractions of the pattern
of atomic displacements $\xi_{{\rm BTO}}$, defined as the difference
between the atomic positions in the relaxed ground states and
the paraelectric reference structure. 
Second, for each frozen fraction of $\xi_{{\rm BTO}}$, the $c$ cell parameter
was relaxed in order to include the strain effects in the internal energy.
Then, the corresponding polarization was approximated by
$P_{\rm BTO}^{0}$=$\frac{\sum_i\xi_{{\rm BTO},i} Z_i^{\ast}}{\Omega}$,
where $\xi_{{\rm BTO},i}$ is the displacement of atom $i$ with respect to its
high symmetric position, Z$_i^{\ast}$ its Born effective charge 
[with the geometry used in the present work, the $(z,z)$ component
of the Born effective charge tensor], 
and $\Omega$ the volume of the unit cell.
The ``0'' in the superscript makes reference to the fact that
this polarization is estimated assuming zero macroscopic electric field.
Finally, the internal energy is plotted as a function of the
zero-field polarization (red curve in Fig.~\ref{wells}), 
and fitted to a polynomial expansion

\begin{equation}
   U_{\rm BTO} (P_{\rm BTO}^{0}) = A_{\rm BTO} P_{\rm BTO}^{0^2} +
                                   B_{\rm BTO} P_{\rm BTO}^{0^4} + 
                                   C_{\rm BTO} P_{\rm BTO}^{0^6} ,
   \label{ubt}
\end{equation}

\noindent where $A_{\rm BTO}$, $B_{\rm BTO}$ and $C_{\rm BTO}$ 
are fitting parameters, 
reported in Table \ref{FittedParametersBulks}. 
The sixth order term was needed to
ensure better agreement of the fit with the first-principles data.

The only way to polarize BaO is within the subspace spanned by its TO modes. 
In the superlattice, the BaO layer is expected to be polarized by BaTiO$_3$
along the $z$ direction and so the evolution of its associated internal 
energy $U_{\rm BO}$ was determined by freezing the pattern of 
displacements of the TO
mode along the $z$ direction given in Table~\ref{ZepsBulk}. As for BaTiO$_3$ 
the out-of-plane cell parameter was relaxed for each amplitude of the frozen
pattern of displacements. Since BaO is not ferroelectric, to polarize it
has an energy cost and the shape of the internal energy corresponds to a
single well (blue curve in Fig.~\ref{wells}) that can be approximated as 

\begin{equation}
   U_{\rm BO} (P_{BO}^{0}) = A_{\rm BO} P_{\rm BO}^{0^{2}} + 
                             B_{\rm BO} P_{\rm BO}^{0^{4}},  
   \label{ubo}
\end{equation}

\noindent where $A_{\rm BO}$ and $B_{\rm BO}$ are also fitting parameters 
reported in 
Table \ref{FittedParametersBulks}. 
Here again, a fourth order term has been included 
to achieve a better agreement in the fit of the energy which slightly deviates
from the harmonic approximation at larger amplitude of polarizations.

\begin{table}[htbp!]
   \begin{center}
      \begin{tabular}{ccccc}
         \hline
         \hline
         $A_{\rm BTO}$     & 
         $B_{\rm BTO}$     & 
         $C_{\rm BTO}$     & 
         $A_{\rm BO}$      & 
         $B_{\rm BO}$      \\
         \hline
         -0.0239           & 
          9.369 10$^{-6}$  & 
         -4.373 10$^{-10}$ & 
          0.122            & 
          1.514 10$^{-4}$  \\
         \hline
         \hline
      \end{tabular}
      \caption{ Values of the parameters,
                used in Eq. (\ref{ubt}) and Eq. (\ref{ubo}),
                that result from a fit of the bulk internal energies
                under zero macroscopic field versus zero-field polarization
                shown in Fig.~\ref{wells}.
                Units required to produce energies in meV when the polarization
                enters in 
                $\mu$C/cm$^{2}$.}
      \label{FittedParametersBulks}
   \end{center}
\end{table}

\subsection{Electrostatic energy cost}

In ferroelectric/incipient-ferroelectric superlattices, 
the incipient ferroelectric layer is expected to become polarized
with nearly the same spontaneous polarization as the ferroelectric layer, 
resulting in a roughly homogeneous polarization through the whole 
structure,\cite{dawber2005} so that $E_{elec}$ in 
Eq. (\ref{Etot1}) will vanish.
Here, in the BaTiO$_3$/BaO superlattices, the BaO layer is 
less polarizable and is no more expected to have necessarily the
same polarization than the BaTiO$_3$ layer.
The polarization misfit gives rise to electric fields and 
produce an additional electrostatic energy cost.
Assuming that both layers are homogeneously polarized, its expression 
can be deduced from the expansion of the total energy of a dielectric
with respect to the electric field $\mathcal{E}$.\cite{sai2002}
For the case of a single dielectric,

\begin{equation}
   E(\mathcal{E}) = E{(\mathcal{E}=0)} -
                    \mathcal{E}P^0     -
                    \frac{1}{2}\epsilon_0\epsilon^{\infty}\mathcal{E}^2.
   \label{eq:expansionsingle}
\end{equation}

\noindent Generalizing Eq. (\ref{eq:expansionsingle}) to the present context,
where we have two dielectrics periodically repeated in the
superlattice, then the zero field energy can be written as

\begin{equation}
   E( \mathcal{E} = 0 ) = m U_{\rm {BTO}} (P_{\rm BTO}^0) 
                        + n U_{\rm{BO}} (P_{\rm BO}^0), 
   \label{equUtot}
\end{equation}

\noindent and the electrostatic energy for the superlattices 
can be written as 

\begin{align}
    & E_{elec}(P_{\rm BTO}^0,P_{\rm BO}^0,
             \mathcal{E}_{\rm BTO},\mathcal{E}_{\rm BO}) = &\nonumber \\
    &    -\Omega_{\rm BTO} \mathcal{E}_{\rm BTO}P_{\rm BTO}^0 
         -\frac{\Omega_{\rm BTO}}{2}\epsilon_0\epsilon^{\infty}_{\rm BTO}
          \mathcal{E}_{\rm BTO}^2 & \nonumber \\
    &   -\Omega_{\rm BO}\mathcal{E}_{\rm BO}P_{\rm BO}^0
         -\frac{\Omega_{\rm BO}}{2}\epsilon_0\epsilon^{\infty}_{\rm BO} 
          \mathcal{E}_{\rm BO}^2, & 
    \label{E_elec1}
\end{align}

\noindent where $\Omega_{\rm BTO}$ and $\Omega_{\rm BO}$ 
are, respectively, the volumes of the BaTiO$_{3}$ and the BaO layers
and can be estimated as 

\begin{subequations}
   \begin{align}
     \Omega_{\rm BTO} & = m c_{\rm {BTO}} S,
     \label{eq:volBTO} 
     \\
     \Omega_{\rm BO} & = n c_{\rm {BO}} S,
     \label{eq:volBO} 
   \end{align}
\end{subequations}

\noindent with $S$ the area of the surface of the unit cell 
[$S = (3.84$ \AA $)^{2}$] and $c_{\rm {BTO}}$ and $c_{\rm {BO}} $
are approximated by $c^0_{\rm {BTO}}$ and $c^0_{\rm {BO}}$.~\cite{note2}

The electric fields inside the BaTiO$_{3}$ layer,
$\mathcal{E}_{\rm BTO}$, and the BaO layer, $\mathcal{E}_{\rm BO}$,
are not independent.
From the short-circuit boundary conditions across the whole 
supercell

\begin{equation}
   m \: c_{\rm{BTO}} \: \mathcal{E}_{\rm BTO} + 
   n \: c_{\rm{BO}} \: \mathcal{E}_{\rm BO} = 0,
   \label{eq:short-circuit}
\end{equation}

\noindent the continuity of the normal component of the displacement field
at the interface

\begin{equation}
   \epsilon_{0} \mathcal{E}_{\rm BTO} + P_{\rm{BTO}} =
   \epsilon_{0} \mathcal{E}_{\rm BO} + P_{\rm{BO}}, 
   \label{eq:continuityd}
\end{equation}

\noindent and the constitutive relations

\begin{subequations}
   \begin{align}
     P_{\rm{BTO}}  & = P_{\rm{BTO}}^{0} + 
                   \epsilon_{0} \chi_{\rm{BTO}}^{\infty} \mathcal{E}_{\rm BTO},
     \label{eq:constitutivepbto} 
     \\
     P_{\rm{BO}}  & = P_{\rm{BO}}^{0} + 
                     \epsilon_{0} \chi_{\rm{BO}}^{\infty} \mathcal{E}_{\rm BO},
     \label{eq:constitutivepbo} 
   \end{align}
\end{subequations}

\noindent we can arrive to expressions for 
$\mathcal{E}_{\rm BTO}$ and $\mathcal{E_{\rm BO}}$,

\begin{subequations}
   \begin{align}
     \mathcal{E}_{\rm BTO}  & = 
        -\frac{n \: c_{\rm BO}  \left(P_{\rm BTO}^{0}-P_{\rm BO}^{0} \right)}
        {\epsilon_{0} \left( n\:c_{\rm BO}\:\epsilon_{\rm BTO}^{\infty} + 
                      m\:c_{\rm BTO}\: \epsilon_{\rm BO}^{\infty} \right)},
     \label{eq:ebto} 
     \\
     \mathcal{E}_{\rm BO}  & = 
        \frac{m \: c_{\rm BTO}  \left(P_{\rm BTO}^{0}-P_{\rm BO}^{0} \right)}
        {\epsilon_{0} \left( n\:c_{\rm BO}\:\epsilon_{\rm BTO}^{\infty} + 
                      m\:c_{\rm BTO}\: \epsilon_{\rm BO}^{\infty} \right)}.
     \label{eq:ebo} 
   \end{align}
\end{subequations}

\noindent Replacing Eqs. (\ref{eq:ebto})-(\ref{eq:ebo}) into 
Eq. (\ref{E_elec1}), then the electrostatic energy reduces to

\begin{equation}
   E_{elec}(n, m, P_{\rm BTO}^{0},P_{\rm BO}^{0}) =
   C(n,m) (P_{\rm BTO}^{0}-P_{\rm BO}^{0})^2,
   \label{eq:E_elec1-bis}
\end{equation}

\noindent where

\begin{align}
  C(n, m) = \frac{S}{2\epsilon_0}\ 
            \frac{n \: m \: c_{{\rm BO}} \: c_{{\rm BTO}}}
             {(m \: c_{{\rm BTO}}\: \epsilon_{\rm BO}^{\infty} + 
             n \: c_{{\rm BO}} \: \epsilon_{\rm BTO}^{\infty})}.
  \label{C_elec}
\end{align}

\noindent Here again, we will approximate $c_{\rm {BTO}}$ and 
$c_{\rm {BO}}$ by $c^0_{\rm {BTO}}$ and $c^0_{\rm {BO}}$.

Combining Eq. (\ref{equUtot}) and Eq. (\ref{eq:E_elec1-bis}), 
the total energy for the electrostatic model for 
(BaTiO$_3$)$_m$/(BaO)$_n$ superlattices reads

\begin{align}
    E(n, m, P_{\rm BTO}^{0},P_{\rm BO}^{0} ) = & 
      m \: U_{\rm BTO}(P_{\rm BTO}^{0}) + n  U_{\rm BO}(P_{\rm BO}^{0})
    \nonumber\\
      & + C(n,m) (P_{\rm BTO}^{0}-P_{\rm BO}^{0})^2.
   \label{Etot}
\end{align}

\subsection{Minimization of the total energy}

For each periodicity $m/n$, the ground state of the superlattice 
can be estimated by minimizing the total energy Eq. (\ref{Etot})
with respect to $P_{\rm BTO}^{0}$ and $P_{\rm BO}^{0}$. 
Then, from the knowledge of the zero-field polarization,
the constitutive relations
[Eqs. (\ref{eq:constitutivepbto})-(\ref{eq:constitutivepbo})], and 
the electric fields [Eqs. (\ref{eq:ebto})-(\ref{eq:ebo})], 
we can infer values for the total polarization
inside each layer.
In Table~\ref{ResultsElecModel} 
we report the results of such minimization for a set of periodicities $n/m$.
It is interesting to see that at a fixed number of BaO layers $n$, 
the model predicts a critical ratio $m/n$ = 3.5 beyond which the
system becomes ferroelectric: 7/2, 14/4 and 21/6.
For all of these ferroelectric states, the minimum of energy is 
reached when the polarization of the BaO layer is about 35\% smaller
than the polarization of the BaTiO$_3$ layer, giving rise to an 
electrostatic energy of the same order of magnitude than the internal
energy. We notice that, considering the internal energy 
alone (\textit{i.e.} assuming $P_{\rm BTO}^{0}$=$P_{\rm BO}^{0}$ 
and E$_{elec}$=0),
we can also predict a critical ratio $m/n\simeq5$ to get a
ferroelectric state (11/2, 21/4 and 31/6). 
This critical ratio coming from the pure internal energies is 
however larger than the critical ratio ($m/n\simeq$3.5) calculated 
with the electrostatic energy, meaning that the quantity of BaTiO$_3$
to get ferroelectricity in the superlattice is lowered by polarizing 
the BaO layer less than the BaTiO$_3$ layer, a situation distinct 
from that usually observed in ferroelectric/incipient-ferroelectric 
superlattices. 
As the polarizability of the insulating materials increases however,
we will tend toward the situation where $P_{\rm BTO}^{0}$=$P_{\rm BO}^{0}$.

\begin{table}[htbp!]
   \begin{center}
      \begin{tabular}{cc r@{.}l r@{.}l r@{.}l r@{.}l r@{.}l}
         \hline
         \hline
         $n$                                            & 
         $m$                                            &
         \multicolumn{2}{c}{$P_{\rm BTO}^{0}$}          &
         \multicolumn{2}{c}{$P_{\rm BO}^{0}$}           & 
         \multicolumn{2}{c}{$U$}                        &
         \multicolumn{2}{c}{$E_{elec}$}                 & 
         \multicolumn{2}{c}{$E_{tot}$}                  \\
         \hline
         2 &  7 & 3&2  & 2&2  &-0&58 &  0&57   &-0&01   \\
         2 &  8 & 9&6  & 6&3  &-6&69 &  5&55   &-1&14   \\
         2 &  9 & 13&0 & 8&4  &-14&95 &  11&03   &-3&92 \\
         2 & 10 &15&4  & 9&9 &-24&77&  16&64  &-8&13    \\
         2 & 11 &17&3  & 11&0 &-35&76&  22&17  &-13&59  \\
        \hline 
         4 & 14 & 3&2 & 2&2 &  -1&17 & 1&14 & -0&03	 \\
         4 & 15 & 7&2 & 4&8 &  -6&74 & 6&03 & -0&71     \\
         4 & 16 & 9&6 & 6&3 &  -13&38 & 11&10 & -2&28   \\
         4 & 17 & 11&4 & 7&5 &  -21&10& 16&43& -4&67    \\
         4 & 18 & 13&0& 8&4 &  -29&90& 22&07& -7&83     \\
        \hline
         6 & 21 & 3&2 & 2&2 &  -1&75 &  1&71 & -0&04    \\
         6 & 22 & 6&2 & 4&1 &  -7&09 &  6&52 & -0&58    \\
         6 & 23 & 8&1 & 5&4 &  -13&19 &  11&48 & -1&71  \\
         6 & 24 & 9&6 & 6&3 &  -20&07&  16&65 & -3&42   \\
         6 & 25 & 10&9 & 7&1 &  -27&79&  22&11& -5&68   \\
        \hline
        \hline
      \end{tabular}
      \caption{ Polarizations $P_{\rm BTO}^{0}$ 
                and $P_{\rm BO}^{0}$ ($\mu$C/cm$^{2}$) 
                minimizing the total energy of Eq. (\ref{Etot})
                for a range of different layer thicknesses $n$ and $m$.
                The corresponding internal, electrostatic and total energies
                (respectively $U$, $E_{elec}$ and $E_{tot}$)
                are also reported. Units of the energies in meV.}
      \label{ResultsElecModel}
   \end{center}
\end{table}

\section{First-principles results}

As a second step, we have performed DFT calculations 
(i) to verify to which extent the predictions of the simple 
electrostatic model are verified, and
(ii) to analyze the eventual existence of instabilities in the 
phonon spectra, that would indicate the existence of
structural phase transitions.

\subsection{Paraelectric reference}

In its paraelectric reference state with space group 
\textit{P4/mmm} (No 123), the in-plane atomic positions in
each layer correspond to the bulk high symmetric positions. 
In the out-of-plane direction a mirror symmetry plane 
is present at the center of each layer. Since in the \textit{z}
direction the atoms are not in their high symmetry positions, 
atomic relaxations take place and give rise to rumpling distortions.
We recover the previous first-principles results of 
Ref.~\onlinecite{junquera2003b}, showing that the rumpling distortions 
are mainly localized at the interface between the layers and rapidly
vanish at the center of thick enough layers,
in good agreement also with the ``locality principle'' discussed
in Ref.~\onlinecite{Stengel-09.c}.
This interface effect in the superlattices involves also a
modification of the interplanar distances between atomic planes 
with respect to those in the bulks, but this effect is also strongly 
located at the interface between the two 
layers.\cite{junquera2003b} As shown in Table~\ref{cell}, 
this feature tends to compress the total out-of-plane cell 
parameter of the multilayer. In spite of the different periodicities 
considered, a constant deviation is calculated between the bulk 
references and the relaxed superlattices confirming the fact that 
the differences with respect to the bulks are mainly located at 
the interface between the two layers.

\begin{table}[htbp!]
   \begin{center}
      \begin{tabular}{ccccc}
         \hline
         \hline
         $n$                 & 
         $m$                 &
         $c^0_{{\rm relaxed}}$ & 
         $c^0_{{\rm bulk}}$    & 
         deviation           \\
         \hline
         2 & 2 & 13.43 & 13.50 & -0.07 \\
         2 & 4 & 21.45 & 21.52 & -0.07 \\
         2 & 6 & 29.46 & 29.54 & -0.08 \\
         2 & 8 & 37.48 & 37.56 & -0.08 \\
         4 & 2 & 18.91 & 18.98 & -0.07 \\
         4 & 4 & 26.93 & 27.00 & -0.07 \\
         4 & 6 & 34.95 & 35.02 & -0.07 \\
         4 & 8 & 42.95 & 43.04 & -0.09 \\
         4 &10 & 50.99 & 51.06 & -0.07 \\
         \hline
         \hline
      \end{tabular}
      \caption{ Out-of-plane cell parameter, $c^0$, for a range of 
                different layer thicknesses $n$ and $m$.
                $c^0_{\rm {relax}}$ represents the perpendicular cell 
                parameter of the paraelectric relaxed case,
                $c^0_{{\rm bulk}}$ stands for the sum of the 
                corresponding bulk unit cell, 
                $c^0_{{\rm bulk}}$ = $m \: c^0_{{\rm BTO}} +
                \frac{n}{2} \: c^0_{{\rm BO}}$. 
                The last column reports the deviation 
                between $c^0_{{\rm bulk}}$ and $c^0_{{\rm relax}}$.
                Units in \AA.}
      \label{cell}
   \end{center}
\end{table}

Since the ferroelectric instabilities are strongly sensitive 
to the volume and pressure, the reduction of the out-of-plane 
thicknesses in the superlattices could strongly influence the
out-of-plane ferroelectric soft mode for the smallest periodicities.
For example, in the 2/2 periodicity, this reduction with respect to the 
bulk reference equals 0.07 \AA,
which corresponds to an equivalent pressure of about 1 GPa.

\subsection{Zone-center instability}

\begin{center}
   \begin{figure}[htbp!]
      \begin{center}
         \includegraphics[width=7.5cm, height=4.17cm, angle=0]{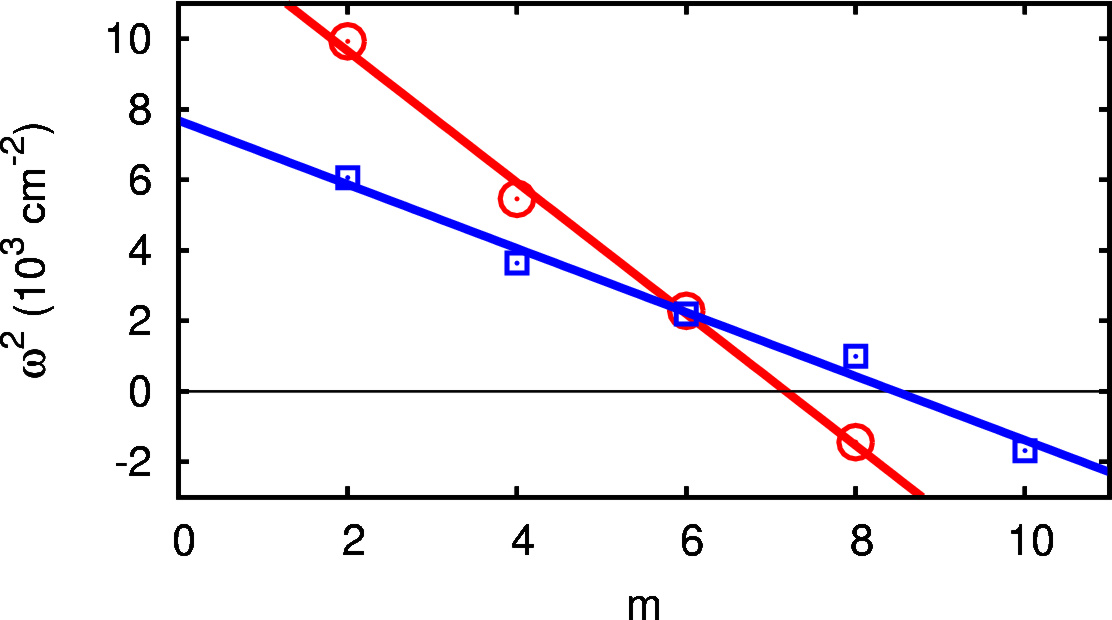}
      \end{center}
      \caption{(Color online) Square of the frequency of the lowest
               zone center out-of-plane transverse optic mode, 
               $\omega^{2}_{\rm TO}$, 
               for $n$=2 (red, circles) and $n$=4 (blue, squares)
               as a function of the BaTiO$_3$ thickness, $m$.
               The phonon frequencies have been computed
               in the reference paraelectric structure. }
      \label{omegaG}
   \end{figure}
\end{center}

For two fixed BaO thicknesses, $n$=2 and $n$=4, the evolution of the 
square of the frequency of the lowest zone center TO mode along the 
out-of-plane direction (FE$_z$), $\omega_{TO}^{2}$, 
calculated in the relaxed paraelectric references, 
is shown in Fig.~\ref{omegaG}. This FE$_z$ mode, that after
condensation will be responsible of the ferroelectric transition, is 
strongly sensitive to the thickness of BaTiO$_3$ layer, 
as expected since  the ferroelectric instability is coming from 
BaTiO$_3$. As shown in Fig.~\ref{omegaG}, $\omega^{2}_{\rm TO}$ 
goes from high values at the smallest $m$ and decreases 
linearly with the increase in the number of BaTiO$_3$ unit cells 
for both $n$=2 and $n$=4 BaO thicknesses.
Beyond a critical $m/n$ periodicity, 7/2 and 9/4, the FE$_z$ mode becomes 
unstable (negative $\omega^2_{\rm TO}$).
This means that, in spite of the non-ferroelectric character of BaO layer, 
the entire supercell can develop a ferroelectric instability when a 
critical ratio between $m/n$ is reached.

This confirms the results predicted with the simple electrostatic model 
Eq. (\ref{Etot}). The agreement is amazingly good for $n=2$ but  the 
critical periodicities at which the ferroelectricity appears do not correspond 
to a constant ratio as inferred from the model:  7/2 and 9/4 from the 
first-principles phonon calculations instead of 7/2 and 14/4  from the 
electrostatic model. Since the electrostatic model is built through the 
{\it bulk} soft mode eigendisplacements and the {\it bulk} Born effective 
charges (to estimate the polarization), we now investigate to which 
extent this constitutes a reasonable approximation.

To highlight the differences between the bulks and the superlattice 
we report in Table~\ref{BECsuperlat} the Born effective charges
of individual atoms at the interface and in the middle of each 
layer of the superlattice in the paraelectric reference structure. 
In the middle of the BaO layer, the amplitude of the $zz$ component 
of the Born effective charges are larger than their bulk values,
a difference that is amplified when $m$ increases and 
reduced when $n$ increases. In the middle of BaTiO$_3$ the 
opposite behaviour is observed, with all the amplitudes of 
the Born effective charges smaller than in the BaTiO$_3$ bulk,
and a trend to reach the bulk values when $m$ increases while they are
reduced when $n$ increases. These global evolutions of the Born
effective charges with $n$ and $m$ in the superlattices can be
related to the natural disposition to recover the 
BaO or BaTiO$_3$  bulk values when
respectively $n$ or $m$ increases. 
The evolution of the Born effective charges of O$_1$, Ti and 
O$_{2/3}$ atoms at the interface are comparable to those in the middle of 
the BaTiO$_3$ layer, but with larger deviations from the bulk amplitudes.
The Born effective charges of the Ba atom at the interface increases
with $m$ and decreases with $n$, but reach values for large $m$ higher
than the BaO or the BaTiO$_3$ bulks. 
The main conclusion that can be drawn from this discussion
is that the Born effective
charges of the atoms in the superlattice are significantly different
than those at the bulk level, mainly for the smallest periodicities $m/n$.

\begin{table}[htbp!]
   \begin{center}
      \begin{tabular}{l r@{.}l r@{.}l r@{.}l r@{.}l r@{.}l r@{.}l r@{.}l r@{.}l r@{.}l}
         \hline
         \hline
                                   &  
         \multicolumn{2}{c}{2/2}   & 
         \multicolumn{2}{c}{4/2}   & 
         \multicolumn{2}{c}{6/2}   & 
         \multicolumn{2}{c}{8/2}   & 
         \multicolumn{2}{c}{2/4}   & 
         \multicolumn{2}{c}{4/4}   &  
         \multicolumn{2}{c}{6/4}   & 
         \multicolumn{2}{c}{8/4}   &
         \multicolumn{2}{c}{Bulk}  \\
         \hline 
         Ba                                    &  
         3&72    &   4&01  &   4&16  &   4&25  &  
         3&14    &   3&38  &   3&51  &   3&61  &  2&78  \\
         O                                     & 
         -3&10   &  -3&34  &  -3&46  &  -3&53  &
         -3&15   &  -3&38  &  -3&53  &  -3&63  &  -2&78  \\
         \hline
         Ba                                    &  
         2&65    &   2&84  &   2&95  &   3&02  &  
         2&50    &   2&69  &   2&79  &   2&87  & \multicolumn{2}{c}{ } \\
         O$_1$                                 &
         -3&75   &  -4&00  &  -4&17  &  -4&26  & 
         -3&53   &  -3&78  &  -3&95  &  -4&06  & \multicolumn{2}{c}{ } \\
         Ti                                    &  
         5&11    &   5&51  &   5&73  &   5&85  &  
         4&83    &   5&20  &   5&43  &   5&58  & \multicolumn{2}{c}{ } \\
         O$_{2/3}$                             &  
         -1&68   &  -1&82  &  -1&88  &  -1&92  & 
         -1&59   &  -1&71  &  -1&78  &  -1&83  & \multicolumn{2}{c}{ } \\
         \hline
         Ba                                    &   
         2&32    &   2&41  &   2&50  &   2&55  &  
         2&20    &   2&28  &   2&38  &   2&44  &  2&77  \\
         O$_{1}$                               & 
         -4&27   &  -4&59  &  -4&78  &  -4&88  & 
         -4&04   &  -4&37  &  -4&55  &  -4&67  & -5&29  \\
         Ti                                    &
         \multicolumn{2}{c}{-}  &  5&88  &  6&13  &  6&27  &
         \multicolumn{2}{c}{-}  &  5&56  &  5&83  &  6&00  &  6&80  \\
         O$_{2/3}$                             & 
         \multicolumn{2}{c}{-}  & -1&86  & -1&93  & -1&97  &
         \multicolumn{2}{c}{-}  & -1&76  & -1&8 3 & -1&88  & -2&14  \\
         \hline
         \hline
      \end{tabular}
      \caption{$zz$ components of the Born effective charges (e) of atoms 
               in the reference paraelectric BaTiO$_3$/BaO superlattices 
               with different periodicities $m/n$. 
               The first two lines refer the Ba and O atoms in the middle 
               of the BaO layer. 
               Lines 3rd to 6th refer to the Ba, Ti and O atoms at the 
               interface.
               Lines 7th to 10th refer to the Ba, Ti and O atoms in the 
               middle of the BaTiO$_3$ layer.
               O$_{1}$ refers to the O in the BaO plane while 
               O$_{2/3}$ stands for the two equivalent oxygens in 
               the TiO$_{2}$ planes.}
      \label{BECsuperlat}
   \end{center}
\end{table}

\begin{center}
   \begin{figure}[htbp!]
      \begin{center}
      \includegraphics[width=7.5cm, height=8.5cm, angle=0]{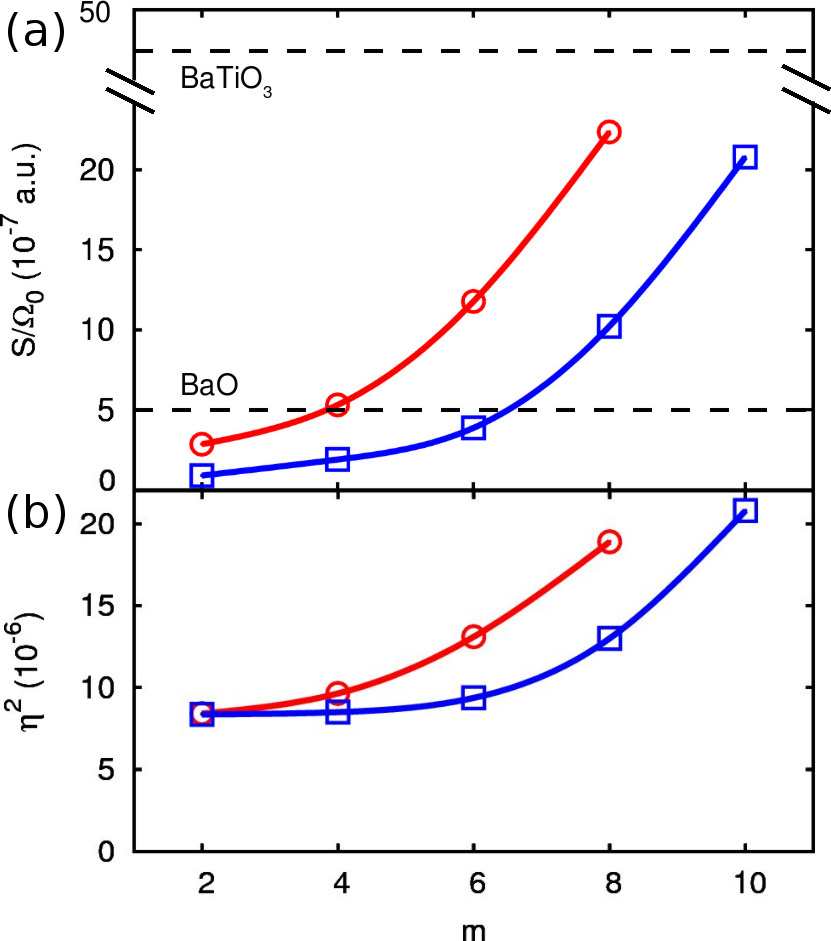}
      \end{center}
      \caption{(Color online) Evolution of the S/$\Omega_0$ ratio (a) 
               and $\eta^2=\sum_{\kappa}\eta^{\ast}_{\kappa}\eta_{\kappa}$ (b)
               of the ferroelectric soft mode for \textit{n}=2 (red, circles)
               and \textit{n}=4 (blue, squares) as a function of the
               BaTiO$_3$ thickness \textit{m}.
               The horizontal dashed lines correspond to the BaTiO$_3$ and
               BaO bulk limit S/$\Omega_0$ values which are respectively
               equal to 43 10$^{-7}$ and 5 10$^{-7}$ a.u.}
      \label{SoverV}
   \end{figure}
\end{center}

Since the Born effective charges of the atoms are modified,
we can expect a modification in the mode polarity of the superlattice
FE$_z$ instability with respect to the bulk. To compare the
evolution of the ferroelectric mode for different
periodicities $m/n$, we report in the Fig.~\ref{SoverV}(a) the
dependence with respect the number of BaTiO$_{3}$ unit cells 
of the ratio between the oscillator strength and the
volume of the unit cell (S/$\Omega_0$) for \textit{n}=2
and \textit{n}=4.
The oscillator strength of one mode is defined as~\cite{gonze1997}

\begin{equation}
   S_{\alpha \beta} =
     \sum_{\kappa \alpha'} Z^{\ast}_{\kappa,\alpha \alpha'} 
          \left( \eta_{\kappa \alpha'}\right)^{\ast}
     \sum_{\kappa' \beta'} Z^{\ast}_{\kappa',\beta \beta'} 
          \eta_{\kappa' \beta'},
   \label{S}
\end{equation}

\noindent where $Z^{\ast}_{\kappa}$ are the individual atomic 
Born effective charge tensors and $\eta$ is the phonon eigendisplacement
vector.

This S/$\Omega_0$ ratio increases strongly with $m$, evolving rapidly
from  values smaller than in bulk BaO at $m$=2 to larger values. 
At the considered thicknesses, S/$\Omega_0$ is much smaller than for
bulk BaTiO$_3$ but we can expect that it will reach the
BaTiO$_3$ bulk value at large enough $m$.
It is clear here that the modification of the S/$\Omega_0$ ratio 
comes either from a modification of the Born effective charge and/or 
from a modification of the eigenvector $\eta$.
As shown in Table~\ref{BECsuperlat}, the Born effective charges 
only tends to increase smoothly with $m$ and cannot be responsible to 
the strong modification of S/$\Omega_0$ shown in Fig.~\ref{SoverV}(a).

In Fig.~\ref{SoverV}(b) we report the evolution of the square of 
the norm of the phonon eigendisplacement vector $\eta^2$ with respect to $m$.
Here, $\eta^2$ increases strongly with $m$, which can be related to a 
strong enhancement  of the ferroelectric eigenvector $\eta$ with 
the BaTiO$_{3}$ thickness. 
Therefore, according to Eq. (\ref{S}), the evolution of the 
S/$\Omega_0$ ratio can be mainly attributed to the modification 
of the ferroelectric eigenvector when $m$ is modified.

In summary, this analysis points out that the simple electrostatic model based on 
the bulk quantities can fail to reproduce the first-principles results because of
a strong modification of the ferroelectric eigenvector in the superlattice  
for the smallest periodicities. This ferroelectric mode is strongly sensitive to the 
thickness of the BaTiO$_3$ layer in the superlattice and, surprisingly does not 
reach the BaTiO$_3$ bulk ferroelectric soft mode, even for the largest $m/n$ 
considered. Compared to other previously studied superlattices like 
SrTiO$_3$/BaTiO$_3$,the modification of the soft-mode pattern could be amplified 
here due to the fact that  the BaTiO$_3$ and the BaO layer are not sharing the same 
perovskite structure. This teaches us that the quantitative agreement of the model 
for $m=2$ could be at least partly fortuitous and that model prediction have to be 
considered with caution.

\subsection{Phonon dispersion curves}
\label{sec:phonondisp}

\begin{figure}[htbp!]
   \begin{center}
      \includegraphics[width=7.5cm, keepaspectratio=true, angle=0]{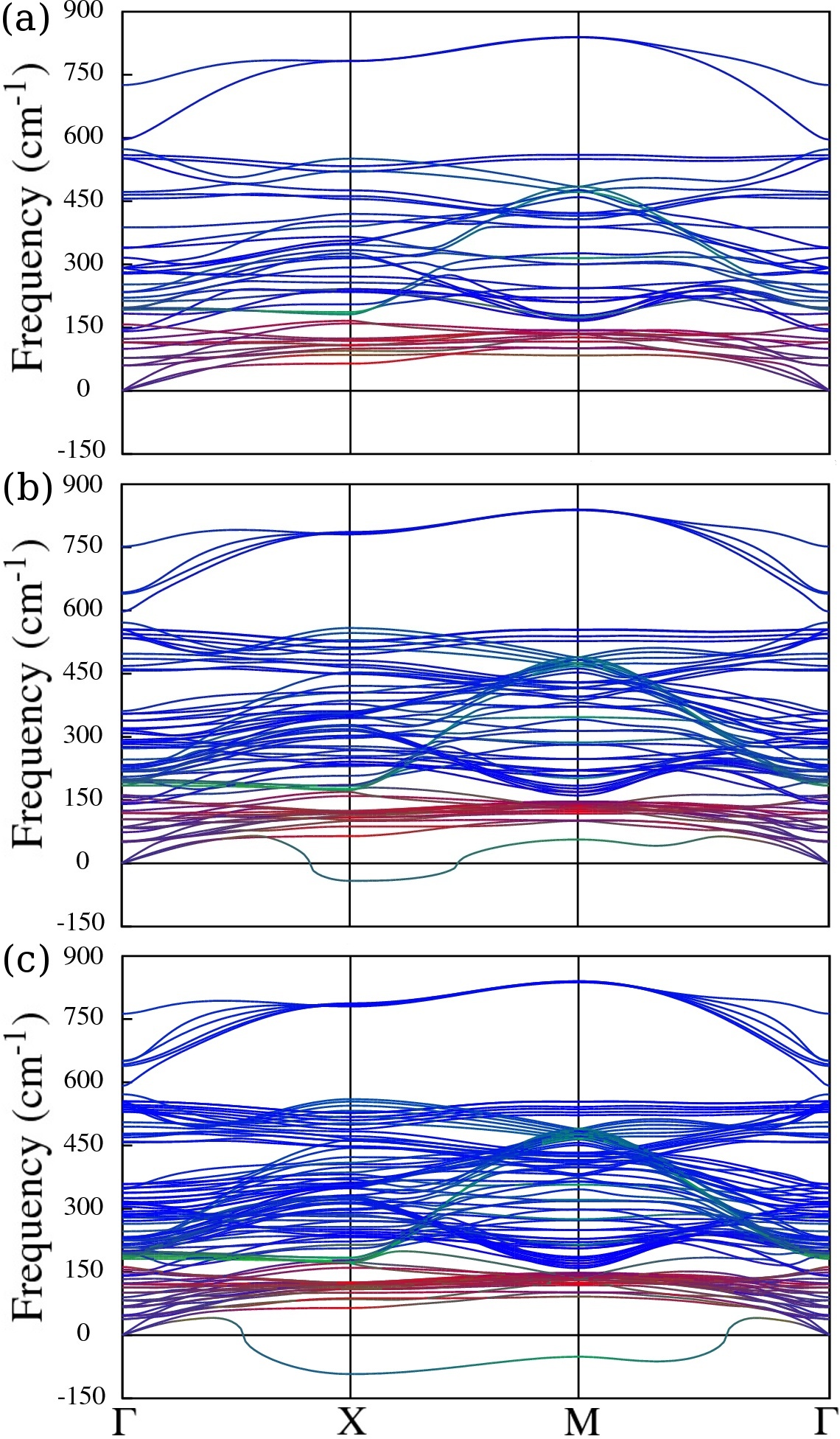} 
      \caption{(Color on line) Phonon dispersion curves 
               for \textit{$m/n$}=2/2 (a), 4/2 (b), and 6/2 (c).
               The color of the different branches are assigned 
               according to the contribution of
               each chemical species to the dynamical matrix eigenvector
               (red for the Ba atom, green for the Ti atom, and blue 
               for the O atom).}
      \label{disp}
   \end{center}
\end{figure}

The phonon dispersion curves between $\Gamma$ (0,0,0), 
X ($\frac{1}{2}$,0,0) and M ($\frac{1}{2}$,$\frac{1}{2}$,0)
points were also calculated, and the results are shown 
in Fig.~\ref{disp} for periodicities 2/2, 4/2 and 6/2.
At the smallest thickness (2/2), no instability is present, 
neither at $\Gamma$, X nor M point.
This confirms that the ground state of the 2/2 multilayer 
is the non-polar \textit{P4/mmm} phase.
However, for the 4/2 periodicity an instability appears at 
the X zone-boundary point with a frequency of 42i cm$^{-1}$,
and its magnitude is amplified for larger $m$, with a calculated 
frequency of 92i cm$^{-1}$ for 6/2.
Moreover, in Fig.~\ref{disp}(b) a low frequency mode is also 
observed at the M zone boundary point for the 4/2 thickness.
This mode becomes unstable for the 6/2 periodicity, but with an 
amplitude smaller than the instability at the X point.
The branch responsible for the M point instability is the same 
than the one responsible for X point instability and its dispersion
is flat between the X and M points while it shifts rapidly to a
positive frequency when going from X or M point to the $\Gamma$ point.
Similar behaviors are also observed for the $m$/4 periodicities. 
An X point instability is also observed for $m\geq$4 and 
its amplitude is also amplified with \textit{m},
with esentially the same imaginary frequencies as in the $n$ = 2 case
(42i cm$^{-1}$ for 4/4 and 91i cm$^{-1}$ for 6/4).
In conclusion, for both thicknesses of the BaO layer,
zone boundary instabilities appear before the 
zone center ferroelectric becomes soft.

\begin{center}
   \begin{figure}[htbp!]
      \begin{center}
 \includegraphics[width=6.5cm,keepaspectratio=true]{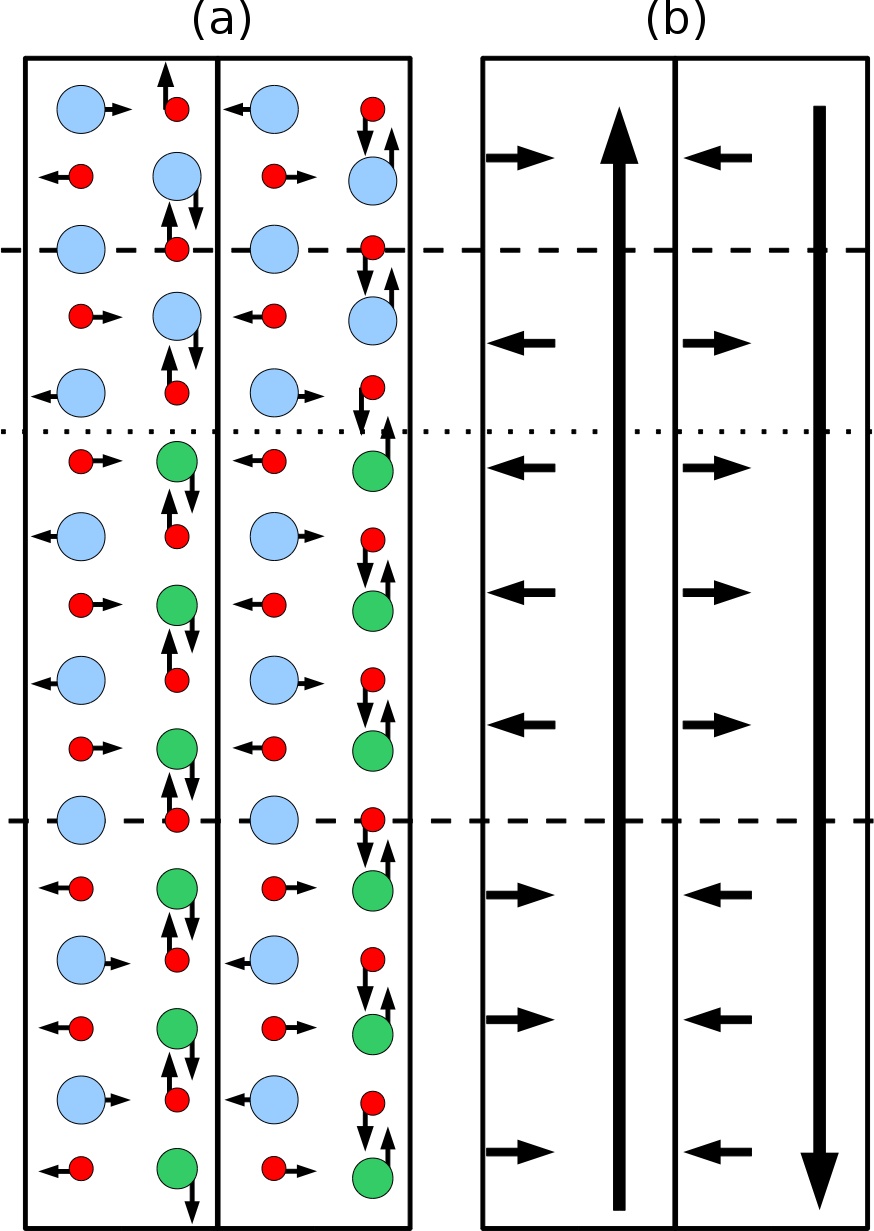}
      \end{center}
      \caption{(Color online) (a) Atomic displacements of the X point unstable 
               mode for the 6/4 thickness. 
               Two consecutive cells along the [100] direction are shown.
               Atoms are represented by balls (O in red, Ti in green and 
               Ba in blue).
               Dashed lines represent the position of the mirror symmetry
               planes of the paraelectric reference structure.
               Dotted lines are the position of the interfaces.
               (b) Schematic representation of the direction of 
               polarization resulting from local polar distortions. }
      \label{afeigen}
   \end{figure}
\end{center}

To clarify the atomic motions related to the X point instability, 
we display a schematic picture of the eigendisplacement pattern in 
Fig.~\ref{afeigen} for the 6/4 periodicity. 
Since the instability is located at the X point, atoms in consecutive 
cells along the [100] direction move out-of-phase. 
As we can see on Fig.~\ref{afeigen}, these eigendisplacements can be
decomposed into two components: 
(i) polar distortions along the \textit{z} direction, only for the
atoms along the Ti--O chains,
and (ii) in-plane motions, only for Ba and O atoms which are along the 
Ba--O chains parallel to the Ti--O chains. 
The amplitudes of these in-plane displacements are however much smaller 
than the out-of-plane displacements, meaning that the Ti--O distortions 
along the $z$ direction dominate the total motions 
(blue-green color of the unstable branch at the X point in 
Fig.~\ref{disp}). Moreover, on the two BaO atomic planes corresponding 
to the two mirror planes of symmetry in the $z$ direction, 
the in-plane motions completely disappear. 
Additionally, we can notice that the polar distortions along the $z$ direction
remain in the BaO layer but only for the atoms aligned with the Ti--O chain, 
in order to preserve the flow of large polarization currents along 
those chains. 

Since in consecutive unit cells along the [100] direction the polar 
displacements are in opposite directions, these X point instabilities 
are related to an \textit{antiferroelectric} distortion. 
To highlight these issues, we report also in Fig.~\ref{afeigen}(b) 
schematic vectors showing the direction of main local dipolar moments. 
From this picture, it is clear that along the $z$ directions we have 
antiferroelectric distortions, with alternating chains of up and down
polarization. For the in-plane Ba and O motions, the atoms move only in the 
[100] direction, giving rise to dipolar moments oriented along the 
$x$ direction. Between the two mirror symmetry planes, 
these in-plane dipolar moments keep the same orientation, 
but it is inverted after crossing these symmetry planes. 
Interestingly, this inversion does not take place at the interface 
of the supercell but in the middle of each layer.

\section{Discussion}

The thickness evolution of the phonon dispersion curves of the 
BaTiO$_3$/BaO superlattice, as discussed in Sec. \ref{sec:phonondisp}, 
can be understood in relationship with those of bulk BaTiO$_3$. 
At the bulk level, BaTiO$_3$ was shown to exhibit a {\it chain-like}
ferroelectric instability in real space, with weak inter-chain interactions, 
yielding ferroelectric modes similarly unstable at
$\Gamma$, X and M points.\cite{ghosez1998,ghosez1999} 
The ferroelectric mode at $\Gamma$ is highly polar. 
In the superlattice, due to the difficulty to polarize BaO,
it will be associated to a huge depolarizing field which will strongly 
harden it. At the opposite, the zone-boundary ferroelectric modes 
at X and M, are globally non-polar. They will therefore not induce
any macroscopic depolarizing field and will so keep a much 
stronger tendency to be unstable than the $\Gamma$ mode.
Consequently, the zone-boundary modes will always be more unstable 
than the $\Gamma$ mode in the superlattice, yielding an anti-ferroelectric 
ground-state. It is only in the limit of thick BaTiO$_3$ layers,
for which the depolarizing field tends to vanish, 
that we will recover the usual ferroelectric ground-state. 
It is also worth to notice that, for $n = 2$ and $4$, the instability at X similarly 
appears at $m=4$ , which corresponds to what was previously 
reported as the approximative length of the correlation volume ($m \approx 4-5$) 
required to induce a ferroelectric instability. \cite{ghosez1998,geneste2008}

The evolution of the ground state of the superlattice 
with the ticknesses of the layers can therefore be summarized as follows:
(i) For the smallest BaTiO$_3$ thicknesses \textit{$m < 4$}, 
the system does not exhibit any ferroelectric instability and the 
supercell remains paraelectric.
(ii) For larger BaTiO$_3$ thicknesses (\textit{$m > 4$}) 
but small ratio \textit{$m/n$}, the ferroelectric instability 
at $\Gamma$ is suppressed by the effect of the depolarizing field; 
there is only an instability at X and the system is antiferroelectric.
(iii) As the ratio \textit{$m/n$} increases, the ferroelectricity 
will progressively expand from X to $\Gamma$ and we can expect 
the formation of ferroelectric domains of increasing sizes.
(iv) Only when $m>>n$, the superlattice will tend to the purely 
ferroelectric ground-state.

We notice that the present BaTiO$_3$/BaO superlattices present similarities 
with Ruddlesden-Popper $A_{n+1}B_nO_{3n+1}$ materials which can 
be viewed as  the stacking of alternating perovskite and rocksalt layers.\cite{Ruddlesden}  
In the latter, however, there is no continuity of the Ti--O chains from one 
perovskite block to the next one. This will prevent the possibility of huge 
polarization currents along the stacking direction as those associated to the
feroelectric mode in BaTiO$_3$/BaO system and could explain why 
Ruddlesden-Popper materials do not develop any tendency to become 
ferroelectric with polarization aligned along the stacking direction.\cite{Nakhmanson2008} 

\section{Conclusions}

In this paper we have studied, from first-principles, superlattices combining $m$ unit cells of 
BaTiO$_3$ with $n$ unit cells of BaO. Although BaTiO$_3$ is ferroelectric at the bulk level, 
we have shown that the superlattice can present an antiferroelectric ground-state and explained
that by the hardening of the ferroelectric mode with respect to the antiferroelectric mode due
to depolarizing field issues.

This behavior is quite general. While epitaxial strain was already previously reported to be able 
to modify the competition between ferroelectric and antiferrodistortive instabilities, we propose
that playing with the depolarizing field in ferroelectric/insulator superlattices is another practical 
way to tune the competition between polar and non-polar instabilites. We have demonstrated here 
the  possibility to engineer an antiferroelectric ground-state in a system combining ferroelectric 
and antiferroelectric instabilities. We anticipate that in systems with competing ferroelectric and
antiferrodistortive instabilities, it could similarly favor an antiferrodistortive ground-state. We 
hope that these results will motivate further studies of feroelectric/insulator superlattices..

\section{Acknowledgement}
This work was supported by the European projects 
CP-FP  228989-2  OxIDes of the Seventh Framework Program, 
the European Multifunctional Institute and the 
Interuniversity Attraction Poles Program (P6/42) - Belgian State - 
Belgian Science Policy. 
JJ acknowledges financial support of the Spanish Ministery of Science and
Innovation through the MICINN Grant. FIS2009-12721-C04-02.
EB also acknowledges FRS-FNRS Belgium.


\begin{thebibliography}{28}
\bibitem[{strainbo()}]{strainbo}
\bibinfo{note}{Experimentally, the lattice constants of BaO
and SrTiO$_3$ fit nearly exactly. The small epitaxial strain of 0.5 \% 
that we report, together with the underestimate of $\omega_{TO}$
(and the related overestimate of the dielectric constant), is an
artefact of the LDA but BaO remains properly described as a paraelectric
so that it does not affect our main conclusions.}
\bibitem[{note2()}]{note2}
\bibinfo{note}{We checked that taking into account the relaxation of $c$ with respect to the polarization does not affect the model predictions.}
\end{thebibliography}

\end{document}